%% file: main.tex
\documentclass[sigconf]{acmart}

\setcopyright{none}
\renewcommand\footnotetextcopyrightpermission[1]{}
\usepackage{microtype}
\usepackage{graphicx}
\usepackage{subcaption}
\usepackage{booktabs}
\usepackage{tabularx}
\usepackage[table]{xcolor}
\usepackage{hyperref}
\usepackage{multirow}
\usepackage{makecell}
\usepackage{amsmath}
\usepackage{amsfonts}
\usepackage{mathtools}
\usepackage{amsthm}
\usepackage{xspace}
\usepackage{enumitem}
\usepackage{algorithm}
\usepackage{algorithmic}
\usepackage[most]{tcolorbox}
\setlist{nosep}

\theoremstyle{plain}

\theoremstyle{definition}

\theoremstyle{remark}

\title[Ensemble-Based Uncertainty Estimation for Code Correctness Estimation]{Ensemble-Based Uncertainty Estimation for Code Correctness Estimation}

\author{Yunxiang Wei}
\email{yunxwei@zju.edu.cn}
\affiliation{
  \institution{Zhejiang University}
  \country{China}
}
\author{Tianlin Li}
\email{tianlin001@buaa.edu.cn}
\affiliation{
  \institution{Beihang University}
  \country{China}
}
\author{Yuwei Zheng}
\email{By2506437@buaa.edu.cn}
\affiliation{
  \institution{Beihang University}
  \country{China}
}
\author{Yanni Dong}
\email{yannidong@outlook.com}
\affiliation{
  \institution{University of Twente}
  \country{Netherlands}
}
\author{Aishan Liu}
\email{liuaishan@buaa.edu.cn}
\affiliation{
  \institution{Beihang University}
  \country{China}
}
\author{Qiang Hu}
\email{qianghu@tju.edu.cn}
\affiliation{
  \institution{Tianjin University}
  \country{China}
}
\author{Xiaoyu Zhang}
\email{xiaoyu.zhang@ntu.edu.sg}
\affiliation{
  \institution{Nanyang Technological University}
  \country{Singapore}
}
\author{Mingfei Cheng}
\email{snowbirds.mf@gmail.com}
\affiliation{
  \institution{Singapore Management University}
  \country{Singapore}
}
\author{Jian Yang}
\email{jiayang@buaa.edu.cn}
\affiliation{
  \institution{Beihang University}
  \country{China}
}

\newcommand{\cas}{\textsc{Cas}\xspace}
\newtcolorbox{rqanswerbox}{
  colback=gray!8,
  colframe=gray!45,
  boxrule=0.4pt,
  arc=2pt,
  left=6pt,
  right=6pt,
  top=4pt,
  bottom=4pt
}

\begin{document}

\begin{abstract}
Large language models (LLMs) have demonstrated remarkable capabilities in generating programs from natural language descriptions, yet ensuring their correctness without an external oracle remains a critical challenge. To solve the challenge, existing methods often rely on uncertainty estimation, measuring the consistency of semantics or execution behaviors across multiple samples generated by a single model. However, we observe that a single model can often converge to a consistent but incorrect solution, rendering such consistency-based proxies ineffective. To address this, we propose \textbf{Ensemble Semantic Entropy (ESE)}, which estimates uncertainty by evaluating the consistency of samples aggregated across an ensemble of models. Experiments on LiveCodeBench demonstrate that ESE correlates more strongly with program correctness than single-model semantic entropy. Notably, in selective generation tasks with strict false-positive rate constraints, ESE improves prediction accuracy by \textbf{53.4\%}. Furthermore, by leveraging ESE as the decision signal, we propose a cascading test-time scaling framework \cas, which maintains performance while reducing FLOPs by \textbf{64.9\%} compared to single-model scaling, offering a new perspective on balancing parameter and inference scaling. 
\end{abstract}



\maketitle

\input{section/intro}
\input{section/preliminaries}
\input{section/method}
\input{section/experiment}
\input{section/tts}
\input{section/related_work}
\input{section/threats}

\section{Conclusion}
\label{sec:conclusion}

Existing uncertainty estimation methods often fail to distinguish between confident errors and correctness due to single-model overconfidence. To address this, we propose \textbf{Ensemble Semantic Entropy (ESE)}, which ensembles diverse models to estimate uncertainty and mitigate false positives. Experiments show that ESE correlates stronger with program correctness, improving prediction accuracy by \textbf{53.4\%} in selective generation task. Furthermore, we propose an ESE-based cascading test-time scaling framework Cas, which reduces FLOPs by \textbf{64.9\%} compared to single-model scaling while preserving performance.

\input{main.bbl}
\input{section/appendix}
\end{document}

%% file: section/intro.tex
\section{Introduction}
\label{sec:introduction}

Large Language Models (LLMs) have shown strong potential in code generation tasks. For instance, Coding assistants such as GitHub Copilot and Cursor can automatically complete code snippets or generate programs from natural language instructions \cite{peng2023the,cursor2026agent}, while agentic tools like Claude Code can further perform complex software engineering tasks, such as repository-level development \cite{anthropic2026claudecode}. As these autonomous tools are increasingly adopted in practical development practice, LLM-generated code is correspondingly becoming deeply integrated into modern software systems.

However, despite their increasing deployment, LLM-generated programs cannot yet be fully trusted. Prior studies show that generated code may still contain hallucination, logical errors, and other defects that lead to incorrect behavior or security-relevant failures in realistic software development settings \cite{liu2023codegeneratedchatgptreally,10.1145/3728894}. While syntax errors and some superficial defects can often be detected by compiler tools, determining whether a syntactically valid program is functionally correct remains difficult. This motivates the need for methods that can predict the correctness of an LLM-generated program before deciding whether to rely on it. A natural approach is to validate candidate programs with existing or self-generated test cases, as seen in recent test-based generation and self-debugging methods \cite{chen2023codet,chen-etal-2025-revisit,10.1007/978-981-95-0014-7_43}. Yet this strategy is often impractical in our setting: comprehensive external tests are rarely available in advance, especially for newly specified problems. Moreover, using LLM-generated tests as an oracle is inherently unreliable, as there is no guarantee that the generated tests themselves correctly capture the program specification, particularly for more challenging tasks \cite{chen-etal-2025-revisit,li2024largelanguagemodelstest,dakhel2023effectivetestgenerationusing,schfer2023empiricalevaluationusinglarge}. 

Recently, utilizing \emph{uncertainty estimation} methods to predict program correctness has gained prominence. Unlike approaches that rely on external oracles, these methods predict correctness from the model's inherent certainty of the generated program. Concretely, these methods typically sample multiple programs from a model and measure the consistency in the semantic space as the uncertainty of the model to the generated programs \cite{kuhn2023semanticuncertaintylinguisticinvariances,nguyen2025semanticentropyboostingllm}. For instance, \cite{ravuri2025eliminatinghallucinationinducederrorsllm}
cluster the sampled programs by their I/O behavior on an LLM-generated test suite and use the size of the largest cluster as a certainty score. \citet{valentin2025incoherenceoraclelessmeasureerror}
formalize the probability that two independently sampled programs produce different I/O behavior as incoherence score and use it as a lower bound on incorrectness .
\citet{sharma2025assessingcorrectnessllmbasedcode}
further use symbolic execution to replace empirical execution behavior on a limited test suite and calculate uncertainty over the induced semantic space. Overall, these approaches are all grounded in the assumption that low uncertainty indicates the output program is likely correct, while high uncertainty is indicative of incorrectness.

However, we observe that although previous uncertainty measures correlated well with incorrectness, as it reflects divergent outputs across multiple samples for a given output, the correlation with correctness is not guaranteed when uncertainty is low (i.e., the model generates programs with consistent semantic behavior). Specifically, a model may exhibit \emph{overconfidence}, causing the generated programs to converge to a consistent but \emph{incorrect} behavior \cite{simhi2025trustmeimwrong,kalai2025languagemodelshallucinate,10.1145/3703155}. We illustrate this with a motivating example in Section~\ref{sec:motivating_example} where a model consistently reproduces identical erroneous logic across syntactically diverse generations. This phenomenon causes existing single-model uncertainty metrics to yield \emph{False Positives} (i.e., estimating low uncertainty for incorrect code), failing to effectively distinguish between ``confident correctness'' and ``confident error''.

\begin{figure*}[t]
    \centering
    \includegraphics[width=0.95\linewidth]{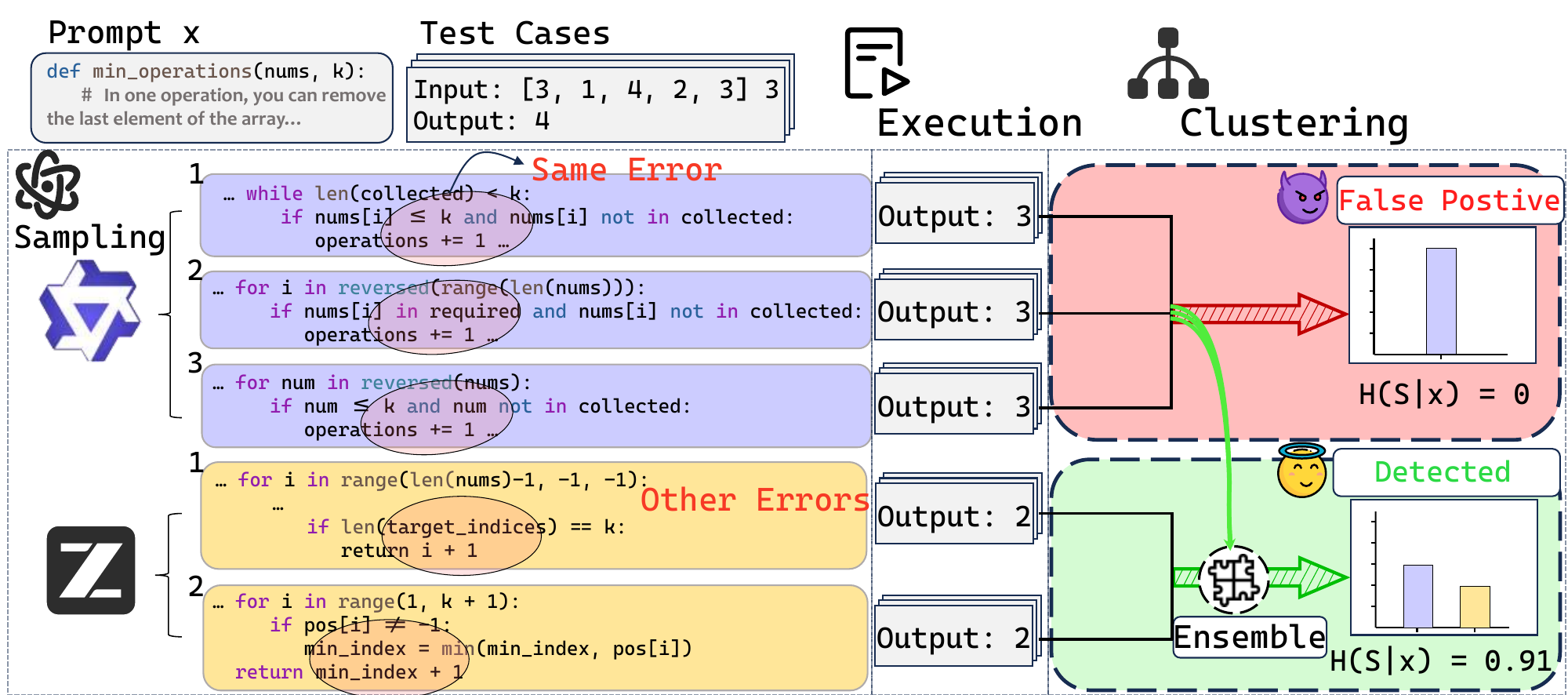}
    \caption{Calculation of Ensemble Semantic Entropy on a motivating example. The problem requires traversing an array \texttt{nums} backwards to collect integers from $1$ to $k$ with minimal steps. \texttt{Qwen3-8B} generates three programs that share the same incorrect semantics (only counting numbers within $[1, k]$), resulting in zero semantic entropy. \texttt{GLM4-9B}, however, generates programs that incorrectly output indices in forward order. By aggregating these five programs, the ensemble reveals the semantic disagreement, resulting in high uncertainty that correctly flags the error, effectively avoiding the false positive.}
    \Description{An example comparing single-model and ensemble semantic entropy for code generation; the ensemble reveals disagreement that catches an incorrect solution.}
    \label{fig:main}
\end{figure*}

To address these limitations, relying on the uncertainty estimation of a single model is inherently insufficient. Our core insight is that while an individual model may confidently converge to a specific error mode, different models typically exhibit distinct failure patterns. Consequently, when multiple models reach a semantic consensus, the credibility of the solution is significantly higher than that derived from single-model self-consistency. Driven by this insight, we propose \textbf{Ensemble Semantic Entropy (ESE)}, a novel uncertainty estimation method that aggregates samples from multiple diverse models and performs clustering and entropy calculation in the semantic space.
By clustering ensembled responses, ESE captures not only the uncertainty within individual models but also the uncertainty arising from distributional differences between models, thereby achieving oracle-free mutual calibration \cite{lakshminarayanan2017simplescalablepredictiveuncertainty,malinin2021uncertaintyestimationautoregressivestructured}.

Experiments on LiveCodeBench \cite{jain2024livecodebenchholisticcontaminationfree} demonstrate that ESE exhibits stronger correlation with program functional correctness than single-model semantic entropy. Notably, in selective generation tasks where model chooses to accept or reject an answer based on uncertainty signal, ESE yields \textbf{53.4\%} improvement in prediction accuracy under strict False Positive Rate (FPR) constraints. Furthermore, we leverage ESE as a reliable decision signal to design \cas, a Cascading Test-Time Scaling framework  \cite{alphacode2,yu2025z1efficienttesttimescaling,samadi2025scalingtesttimecomputeachieve,li2025stesttimescaling}. This framework dynamically evaluates generation quality, allowing smaller models to resolve problems where uncertainty is low and escalating to larger models only when uncertainty is high. Results show that compared to standard Best-of-$N$ scaling, our framework preserves code generation performance while reducing FLOPs consumption by \textbf{64.9\%}, offering a new perspective on balancing Parameter Scaling and Inference Scaling.

Our primary contributions are summarized as follows:
\begin{itemize}
\item{We propose Ensemble Semantic Entropy (ESE), which effectively quantifies uncertainty in code generation by aggregating outputs from diverse models within the program semantic space.}
\item{We validate the ability of ESE to predict program correctness in various experiments, demonstrating its stronger correlation with correctness and better performance in selective generation tasks.}
\item{We introduce an ESE-based Cascading Test-Time Scaling Framework \cas, enabling dynamic resource allocation that significantly reduces inference costs while maintaining generation performance.}
\end{itemize}

%% file: section/preliminaries.tex
\section{Preliminaries}
\label{sec:preliminaries}

% In this section, we formalize the problem of predicting program correctness in code generation. We then revisit standard Bayesian uncertainty and introduce \textit{Semantic Entropy} by defining program equivalence through execution semantics. Finally, we provide a motivating example to illustrate the false positive problem of single-model semantic entropy.

\subsection{Problem Setup}
\label{sec:problem_setup}

We define the code generation task as generating a program $y$ given a prompt $x$. A Code LLM parameterized by $\theta$ induces a conditional distribution $p(y \mid x, \theta)$ over the space of programs $\mathcal{Y}$.

For the generated program, incorrectness can arise from two sources. The first is \emph{syntax error}, where the generated code cannot be parsed or compiled. Such errors can typically be identified with compilers and are not the focus of this work. The second is \emph{functional incorrectness}, where a syntactically valid program executes but does not satisfy the intended specification. Since our goal is to estimate uncertainty for correctness prediction beyond simple syntax checking, in the remainder of this paper we restrict $\mathcal{Y}$ to the space of syntactically valid programs.

For a given specification $x$, there exists a target correct function $f^\star_x$. A generated program $y \in \mathcal{Y}$ is considered correct if and only if it functionally matches $f^\star_x$, regardless of its implementation details. In this work, we focus on code generation tasks with a unique target behavior for each specification, excluding open-ended problems that admit multiple functionally distinct correct solutions.

Our primary goal is to employ \textit{uncertainty estimation} to predict program correctness. Concretely, for a syntactically valid generated program $y$, we aim to derive an uncertainty score $\mathcal{U}(x, y)$ that correlates with the probability that $y$ is functionally incorrect. A high uncertainty score should therefore indicate a high risk that the generated program fails to match the target function $f^\star_x$.
  
\subsection{Bayesian Uncertainty Quantification}
\label{sec:bayesian_uq}

From a Bayesian perspective, we treat the model parameters $\theta$ as a random variable. Given a training dataset $\mathcal{D} = \{(x_i, y_i)\}_{i=1}^N$, training yields a posterior distribution $p(\theta \mid \mathcal{D})$. For a input $x$, the predictive distribution is obtained by marginalizing over the posterior:
\begin{equation}
    p(y \mid x, \mathcal{D}) = \int p(y \mid x, \theta) p(\theta \mid \mathcal{D}) \, \mathrm{d}\theta.
    \label{eq:predictive_posterior}
\end{equation}
The total predictive uncertainty is classically quantified using the Shannon entropy $H$ of this predictive distribution. Let $Y$ denote the random variable for the output sequence:
\begin{equation}
    H[Y \mid x, \mathcal{D}] = - \sum_{y \in \mathcal{Y}} p(y \mid x, \mathcal{D}) \log p(y \mid x, \mathcal{D}).
    \label{eq:total_entropy}
\end{equation}
% This quantity captures the total uncertainty inherent in the prediction given the observed data $\mathcal{D}$, encompassing both data noise (aleatoric uncertainty) and model ignorance (epistemic uncertainty).
\vspace{-0.1cm}
\subsection{Semantic Entropy}
\label{sec:semantic_entropy}

Based on the information-theoretic decomposition of uncertainty \cite{kendall2017uncertaintiesneedbayesiandeep}, the total entropy can be expressed as:
\begin{equation}
    H[Y \mid x, \mathcal{D}] = \underbrace{\mathbb{E}_{\theta \sim p(\theta \mid \mathcal{D})} \big[ H[Y \mid x, \theta] \big]}_{\text{Aleatoric Uncertainty}} + \underbrace{I(Y; \Theta \mid x, \mathcal{D})}_{\text{Epistemic Uncertainty}},
    \label{eq:entropy_decomposition}
\end{equation}
where $\Theta$ is the random variable representing model parameters. 

Standard uncertainty methods typically estimate these quantities in the token space $\mathcal{Y}$. While in code generation tasks, many distinct sequences can implement the same algorithm. Consequently, the standard aleatoric term $H[Y \mid x, \theta]$ tends to overestimate uncertainty by capturing trivial lexical variations rather than meaningful semantic ambiguity. Therefore uncertainty over \textit{functional semantics} is often more relevant than uncertainty over specific token sequences.

Semantic Entropy (SE) \cite{kuhn2023semanticuncertaintylinguisticinvariances} shifts the focus from raw tokens to semantic concepts via clustering samples first. Let $\phi: \mathcal{Y} \to \mathcal{S}$ be a semantic mapping function such that sequences $y$ sharing the same meaning are grouped into a semantic cluster $s \in \mathcal{S}$ (i.e., $\phi(y) = s$). In practice, $\phi$ can be defined by execution based equivalence, which mapped programs into the same cluster if they produced identical outputs on a test set $\mathcal{T}$. Let a random variable $S$ over the semantic space $\mathcal{S}$. The probability of a semantic cluster $s$ under a fixed model $\theta$ is given by aggregating the probabilities of all samples belonging to that cluster:
\begin{equation}
    p(s \mid x, \theta) = \sum_{y: \phi(y)=s} p(y \mid x, \theta).
\end{equation}
The Semantic Entropy for a single model $\theta$ is defined as the entropy of the semantic variable $S$:
\begin{equation}
    H_{\mathrm{SE}}(x; \theta) \coloneqq - \sum_{s \in \mathcal{S}} p(s \mid x, \theta) \log p(s \mid x, \theta).
    \label{eq:se_def}
\end{equation}

% \subsection{Program Semantic Equivalence}
% \label{sec:code_equivalence}
% In practice, the semantic mapping $\phi$ is realized by establishing an equivalence relation $\sim$ on the program space (e.g., mapping a program to its equivalence class) \cite{sharma2025assessingcorrectnessllmbasedcode,ravuri2025eliminatinghallucinationinducederrorsllm}. For code generation, this is defined as \textit{functional equivalence}. Formally, viewing a program $y$ as a function $f_y: \mathcal{I} \to \mathcal{O}$, two programs $y$ and $y'$ are equivalent if $f_y(i) = f_{y'}(i)$ for all $i \in \mathcal{I}$.

% As exact verification of functional equivalence is generally undecidable, existing approaches typically approximate this relation using a set of test inputs $\mathcal{T} = \{t_1, \dots, t_K\}$. By executing a program $y$ on $\mathcal{T}$, we obtain an output vector $\mathbf{o}_y = [f_y(t_1), \dots, f_y(t_K)]$. The equivalence is then determined by these execution traces: $y \sim y' \text{ iff. } \mathbf{o}_y = \mathbf{o}_{y'}$. Here, the semantic space $\mathcal{S}$ is defined as the space of output vectors. Accordingly, all semantic uncertainty measures in the remainder of the paper are defined with respect to this test-induced semantic partition rather than exact functional equivalence. Intuitively, a higher semantic uncertainty indicates that the generated programs are dispersed across multiple distinct semantic clusters, thereby suggesting a higher likelihood of error. 

\section{Motivating Example}
\label{sec:motivating_example}
\begin{figure}[t]
    \centering
    \includegraphics[width=0.5\textwidth]{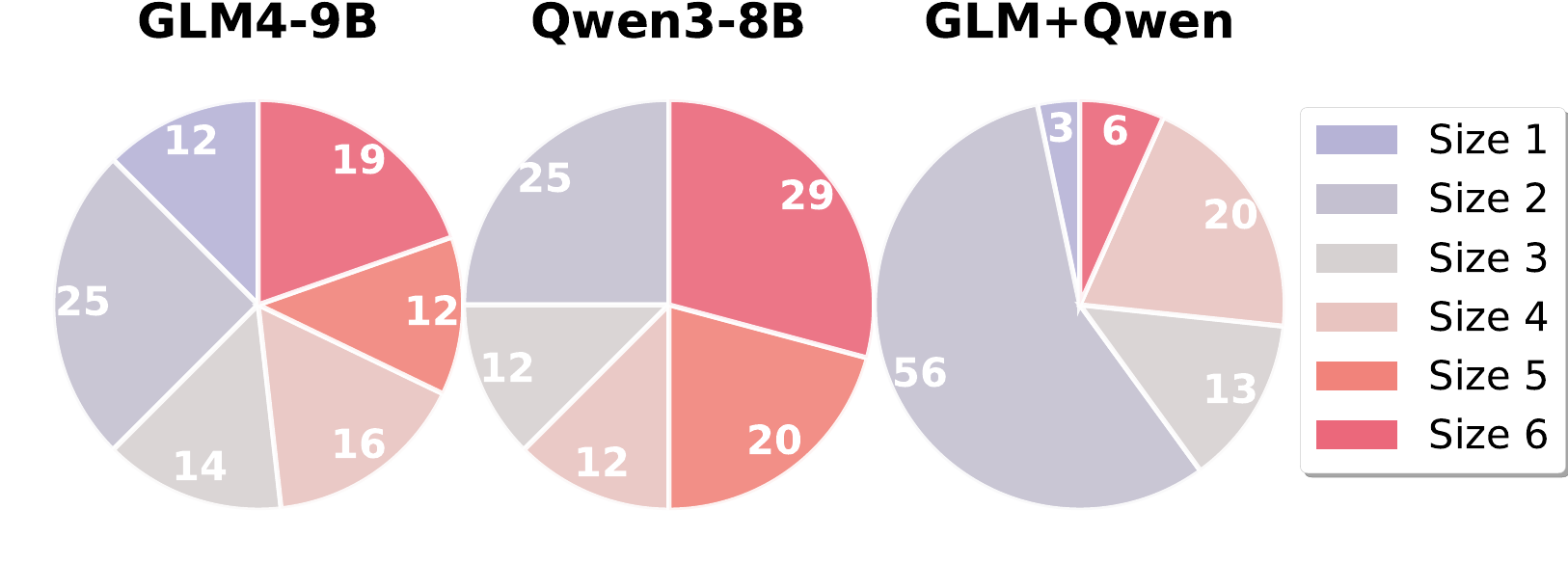}
    \caption{Comparison of the distribution of the largest cluster sizes in incorrect problems. Single models (GLM4-9B, Qwen3-8B) frequently show high consistency in incorrect answers, whereas the ensemble substantially reduces this spurious consistency.}
    \label{fig:cluster_size}
    % \vspace{-0.6cm}
\end{figure}
The premise of using standard SE for correctness prediction is that correctness usually converges to a single semantic behavior, while errors are diverse. However, this assumption can fail when a model exhibits \textbf{overconfidence}, namely, when it repeatedly produces semantically consistent but functionally incorrect programs.

Figure~\ref{fig:main} illustrates a concrete example of this phenomenon. The problem requires traversing an array \texttt{nums} backwards to build a set. The \texttt{Qwen3-8B} model generates three programs that, while implementing the backward traversal differently, share the exact same semantics with same error (missing the final element in total steps count). Since the three programs are functionally equivalent, they are mapped to the same semantic cluster. Thus the single-model SE is 0, falsely indicating that the model is confident about the correctness of the generated programs.

% Figure~\ref{fig:main} illustrates a concrete example of this 
% phenomenon. The problem requires traversing an array \texttt{nums} 
% backwards. In each step, the number at the current position is added 
% to a set, and the traversal continues until the set contains all 
% integers from $1$ to $k$, with the goal of minimizing steps. For the 
% given test cases, operations should be performed on the last four 
% numbers so the output should be 4.
% The \texttt{Qwen3-8B} model generates three programs that, while 
% implementing the backward traversal differently, share the exact same 
% semantics and commit the same error: they only count an operation if 
% the current number falls within the range $[1, k]$, thereby missing 
% the operation on the number $4$.
% Since the three programs are functionally equivalent, they are mapped 
% to the same semantic cluster. Thus the semantic entropy is 0, falsely 
% indicating that the model is confident about the correctness of the 
% generated programs.

\textbf{Overconfidence is prevalent in incorrect generations.}
To quantify the prevalence of this issue, we analyzed the behavior on the LiveCodeBench easy split as shown in Figure~\ref{fig:cluster_size}. We tested the \texttt{GLM4-9B} and \texttt{Qwen3-8B} models by generating 6 programs per problem and using the problem's private tests as a comprehensive test $\mathcal{T}$ for execution based clustering. Programs are grouped into the same cluster if they procuded identical outputs on all tests in $\mathcal{T}$. Focusing on problems where at least three programs were incorrect, we measured the size of the largest cluster. The results show that for \texttt{GLM4-9B} and \texttt{Qwen3-8B}, 31\% and 49\% of these problems, respectively, exhibited a dominant incorrect cluster with a size greater than or equal to 5. In these instances where models are confidently wrong, single-model SE fails to distinguish them from correct cases because the estimated uncertainty remains extremely low.

\textbf{Different models do not share error patterns.}
However, we find that such concentrated error patterns are not typically shared across different models. We constructed an ensemble by sampling 3 programs from \texttt{GLM4-9B} and 3 from \texttt{Qwen3-8B}, forming a set of 6 programs for behavioral clustering. Under the same setting, the proportion of problems with a dominant cluster larger than 5 samples was dramatically reduced to 6\%.

This phenomenon can be understood by applying the uncertainty decomposition in Eq.~\eqref{eq:entropy_decomposition} to the semantic variable $S=\phi(Y)$: 
\begin{equation}
    H[S \mid x, \mathcal{D}] = \mathbb{E}_{\theta \sim p(\theta \mid \mathcal{D})}\!\left[ H[S \mid x, \theta] \right] + I(S;\Theta \mid x, \mathcal{D}).
\end{equation}
The standard SE in Eq.~\eqref{eq:se_def} evaluates $H[S \mid x, \theta]$ at a single fixed model and therefore captures only within-model semantic dispersion. When a model is internally consistent yet concentrates on an incorrect semantic cluster, this quantity can remain low even though the generated program is wrong. In this sense, single-model SE reflects self-consistency rather than correctness \cite{park2026efficientsemanticuncertaintyquantification}. To estimate the full uncertainty, the missing component is the cross-model disagreement term $I(S;\Theta \mid x, \mathcal{D})$, namely semantic epistemic uncertainty. Since different models often do not share the same error pattern, aggregating their semantic distributions may recover part of the missing epistemic uncertainty through disagreement between models on the incorrect programs. This observation motivates estimating semantic uncertainty from an ensemble of diverse models, which we formalize in Section~\ref{sec:method}.

% This phenomenon can be understood through the uncertainty decomposition in Eq.~\eqref{eq:entropy_decomposition}. The total uncertainty contains both an aleatoric term and an epistemic term. In contrast, the standard SE in Eq.~\eqref{eq:se_def} is computed under fixed parameters $\theta$, which act as a point estimate of the posterior. As a result, single-model SE only measures the dispersion of semantic behaviors induced by that model and does not capture the epistemic uncertainty associated with variation in model parameters. When a model confidently concentrates its probability mass on one incorrect semantic cluster, the resulting entropy remains low even though the prediction is wrong. In this sense, single-model SE reflects self-consistency rather than calibration to the unknown true behavior \cite{park2026efficientsemanticuncertaintyquantification}. Since different models often do not share the same error pattern, aggregating their semantic distributions may recover part of the missing epistemic uncertainty through cross-model disagreement. This observation motivates estimating semantic uncertainty from an ensemble of diverse models, which we formalize in Section~\ref{sec:method}.

%% file: section/method.tex
\section{Method}
\label{sec:method}

In this section, we present our method for predicting program correctness from \textbf{Ensemble Semantic Entropy (ESE)}. Motivated by the analysis in Section~\ref{sec:motivating_example}, our key idea is to estimate uncertainty from multiple models rather than from a single model alone, so as to capture not only within-model semantic dispersion but also cross-model disagreement.

We formulate entropy-based uncertainty estimation for code generation in three stages: (i) \textbf{semantic clustering}, which instantiates the semantic mapping for code generation; (ii) \textbf{entropy calculation}, which computes an uncertainty score from model-specific semantic distributions; and (iii) \textbf{correctness prediction}, which uses the uncertainty score to decide whether the generated programs are likely to be correct.

Correspondingly, we first specify how programs are mapped to semantic classes through execution-based functional equivalence, then compute uncertainty using ESE or its black-box variant EDSE, and finally apply a threshold to predict program correctness.

\subsection{Semantic Clustering}
\label{sec:semantic_clustering}

In Section~\ref{sec:semantic_entropy}, Semantic Entropy is defined on a semantic space $\mathcal{S}$ through a mapping $\phi:\mathcal{Y}\to\mathcal{S}$. Here, we instantiate $\phi$ for code generation via functional equivalence.

Let $\mathcal{M} = \{\theta^{(\ell)}\}_{\ell=1}^{L}$ denote an ensemble of $L$ code generation models. For code generation, semantic equivalence is defined by program behavior. Formally, viewing a program $y$ as a function $f_y: \mathcal{I} \to \mathcal{O}$, two programs $y$ and $y'$ are functionally equivalent if $f_y(i) = f_{y'}(i), \forall i \in \mathcal{I}$.
Accordingly, $\phi$ maps all functionally equivalent programs to the same semantic class.

As exact verification of functional equivalence is generally undecidable, we approximate this relation using a problem-specific set of test inputs $\mathcal{T}_x = \{t_1, \dots, t_{R_x}\}$. By executing a program $y$ on $\mathcal{T}_x$, we obtain an output vector $\mathbf{o}_x(y) = [f_y(t_1), \dots, f_y(t_{R_x})]$.
We then instantiate the semantic mapping by this execution behavior, so that $\phi(y) = \mathbf{o}_x(y)$.
Under this approximation, two programs are assigned to the same semantic class if and only if they produce identical execution outputs on $\mathcal{T}_x$.
% $y \sim y' \text{ iff. } \phi(y)=\phi(y')$.
Here, the semantic space $\mathcal{S}$ is given by the space of possible execution output vectors on $\mathcal{T}_x$.

% This execution-based instantiation provides a shared semantic representation for all models in the ensemble. When likelihoods are available, it allows each model's program distribution to be aggregated into a semantic distribution as in Section~\ref{sec:semantic_entropy}; when only finite samples are available, the same mapping is applied to sampled programs to obtain an empirical approximation.

\subsection{Ensemble Semantic Entropy (ESE)}
\label{sec:ese}

Given the semantic mapping $\phi$, each model $\theta^{(\ell)}$ defines a semantic distribution over semantic classes:
\begin{equation}
    p(s \mid x, \theta^{(\ell)}) = \sum_{y:\phi(y)=s} p(y \mid x, \theta^{(\ell)}).
\end{equation}
Single-model Semantic Entropy (SE) computes uncertainty from one such distribution. In contrast, we first aggregates the semantic distributions from all models in the ensemble and then measures the entropy of the aggregated distribution.

We define the \textbf{Ensemble Semantic Probability} for a semantic class $s$ as
\begin{equation}
    p(s \mid x, \mathcal{M}) = \frac{1}{L} \sum_{\ell=1}^{L} p(s \mid x, \theta^{(\ell)}).
    \label{eq:ese_prob}
\end{equation}
where we use uniform weights over the ensemble members. The \textbf{Ensemble Semantic Entropy (ESE)} is then defined as the entropy of this aggregated semantic distribution:
\begin{equation}
    \boxed{
    H_{\mathrm{ESE}}(x) \coloneqq - \sum_{s \in \mathcal{S}} p(s \mid x, \mathcal{M}) \log p(s \mid x, \mathcal{M}).
    }
    \label{eq:ese_def}
\end{equation}

This definition also admits the following decomposition:
\begin{equation}
    H_{\mathrm{ESE}}(x)
    =
    \frac{1}{L}\sum_{\ell=1}^{L} H_{\mathrm{SE}}(x; \theta^{(\ell)})
    +
    \operatorname{JSD}\!\left(
        p(\cdot \mid x,\theta^{(1)}), \ldots, p(\cdot \mid x,\theta^{(L)})
    \right),
\end{equation}
where $\operatorname{JSD}$ denotes the Jensen--Shannon divergence under uniform mixture weights, taken over the semantic distributions on $\mathcal{S}$. Hence, ESE consists of two parts: the average within-model semantic entropy and a cross-model disagreement term. The first term captures semantic dispersion within each individual model, whereas the second term captures how much the models disagree in semantic space.

This decomposition can explain why ESE is better aligned with the motivating example in Section~\ref{sec:motivating_example}. If a single model is internally consistent yet concentrates on one incorrect semantic cluster, its semantic entropy can remain low. However, if different models concentrate on different incorrect behaviors, the disagreement term increases, and the resulting ESE becomes high. In this sense, ESE extends single-model SE by incorporating a practical proxy for semantic epistemic uncertainty.

\subsection{Ensemble Discrete Semantic Entropy (EDSE)}
\label{sec:ens_dse}

Computing $H_{\mathrm{ESE}}$ typically requires access to token-level likelihoods $p(y\mid x,\theta)$ (the \emph{gray-box} setting), which is often infeasible for closed-source LLMs. To address this, existing work suggests a fully \emph{black-box} variant, which is also valid in our framework \cite{Farquhar2024}.

For each model $\theta^{(\ell)}$, let $\{y^{(\ell)}_k\}_{k=1}^{K}$ denote its $K$ sampled programs. After mapping each sample to its semantic class, we estimate the model-specific semantic distribution by empirical frequency:
\begin{equation}
    \hat{p}(s \mid x, \theta^{(\ell)}) = \frac{1}{K} \sum_{k=1}^{K} \mathbb{I}[\phi(y^{(\ell)}_k) = s],
\end{equation}
where $\mathbb{I}[\cdot]$ is the indicator function. Analogously to Eq.~\eqref{eq:ese_prob}, we aggregate these empirical semantic distributions across the ensemble:
\begin{equation}
    \bar{p}_{\mathrm{DSE}}(s \mid x) = \frac{1}{L} \sum_{\ell=1}^{L} \hat{p}(s \mid x, \theta^{(\ell)}).
    \label{eq:dse_aggregated}
\end{equation}
The \textbf{Ensemble Discrete Semantic Entropy (EDSE)} is then defined as
\begin{equation}
    \boxed{
    H_{\mathrm{EDSE}}(x) \coloneqq - \sum_{s \in \mathcal{S}} \bar{p}_{\mathrm{DSE}}(s \mid x) \log \bar{p}_{\mathrm{DSE}}(s \mid x).
    }
    \label{eq:dse_ens_def}
\end{equation}

% EDSE preserves the central idea of ESE, namely, to estimate uncertainty from the semantic distribution induced by an ensemble of models rather than from a single model alone. The difference is that EDSE replaces model likelihoods with sample frequencies, making it directly applicable in black-box settings.

% \subsection{Predicting Correctness via Uncertainty}
% \label{sec:predict_correctness}

Once the uncertainty score has been computed, we use this score to predict whether the generated programs are likely to be correct. Let $u(x)$ denote the uncertainty score for problem $x$, instantiated by either $H_{\mathrm{ESE}}(x)$ or $H_{\mathrm{EDSE}}(x)$. Since higher semantic uncertainty indicates greater dispersion or disagreement among sampled programs, we treat $u(x)$ as a proxy for the risk of functional incorrectness.

We adopt a threshold-based decision rule. Given a threshold $\tau$, we predict that the sampled programs for $x$ are reliable if $u(x) \le \tau$, and unreliable otherwise. This defines a binary predictor that accepts low-uncertainty cases and rejects high-uncertainty cases. In later experiments, we evaluate how well this uncertainty-based decision rule supports correctness prediction.

%% file: section/experiment.tex
\section{Experiments}
\label{sec:experiments}

In this section, we evaluate ESE as an uncertainty signal for predicting program correctness in code generation task. To this end, we study the following research questions:
\begin{itemize}
    \item \textbf{RQ1:} How strongly does ESE correlate with program correctness compared with single-model entropy-based uncertainty methods?
    \item \textbf{RQ2:} How do different semantic clustering methods affect the correlation between ESE and program correctness?
    \item \textbf{RQ3:} How accurately does ESE support selective generation with abstention under strict false-positive constraints?
\end{itemize}

\input{table/auc_detailed}

\subsection{Experimental Setup}

We conduct experiments on \textbf{LiveCode\allowbreak Bench}~\cite{jain2024livecodebenchholisticcontaminationfree} (Release v2), which consists of recent contest problems (e.g., AtCoder, Codeforces) equipped with rigorous, hidden test cases.

Following the formulation in Section~\ref{sec:method}, we compute uncertainty by first clustering sampled programs in semantic space and then applying an entropy calculation method to the induced distribution. Unless otherwise stated, we use the behavior-based clustering method introduced in Section~\ref{sec:semantic_clustering}, named \textit{Func}, which groups programs by their execution outputs on a problem-specific generated test set $\mathcal{T}^{gen}_x$; in our experiments, $\mathcal{T}^{gen}_x$ is pre-generated by \texttt{Qwen3-Coder-30B}. Under this default clustering, we compare five uncertainty scores: \textbf{Predictive Entropy (PE)}, \textbf{Semantic Entropy (SE)}, \textbf{Discrete Semantic Entropy (DSE)}, and our ensemble variants \textbf{ESE} and \textbf{EDSE}.

For each problem $x$ we consider $M=12$ sampled programs. In the single-model setting, all samples are drawn from one model. In the ensemble setting, the samples are evenly drawn from two models and merged before uncertainty computation.

% for each problem $x$ we consider $M=12$ sampled programs. In the single-model setting, all samples are drawn from one model. In the ensemble setting, the samples are evenly drawn from two models and merged before clustering and entropy computation. We consider five uncertainty scores: \textbf{Predictive Entropy (PE)}, \textbf{Semantic Entropy (SE)}, \textbf{Discrete Semantic Entropy (DSE)}, and our ensemble variants \textbf{ESE} and \textbf{EDSE}.

We evaluate models from three families: \texttt{GLM4-9B}, the \texttt{Qwen3} series (\texttt{Qwen3-8B}, \texttt{Qwen3-14B}, \texttt{Qwen3-Coder-30B-A3B}, \texttt{Qwen3-Coder\allowbreak-485B-A35B}), and the \texttt{gpt-oss} series (\texttt{gpt-oss-20B}, \texttt{gpt-oss-120B}) \cite{yang2025qwen3technicalreport,openai2025gptoss120bgptoss20bmodel,glm2024chatglmfamilylargelanguage}. To investigate the effect of ensemble models on uncertainty calibration, we design two-model ensembles spanning both cross-family and within-family pairings. The cross-family settings probe whether model diversity provides a meaningful calibration signal, while the within-family settings test whether the gain persists when the paired models are likely to make more correlated errors.
Specifically, for the smaller models \texttt{Qwen3-8B} and \texttt{gpt-oss-20B}, we pair each with a parameter-comparable model from a different family (respectively, \texttt{GLM4-9B} and \texttt{Qwen3-Coder-30B}).
For mid-scale models (\texttt{Qwen3-14B}, \texttt{Qwen3-Coder-30B}, and \texttt{gpt-oss\allowbreak-120B}), we pair each with the smaller model from the same family to probe within-family ensembling (e.g., \texttt{Qwen3-14B} paired with \texttt{Qwen3-8B}).
Finally, we also explore ensembling two strong large-scale models, \texttt{Qwen3-Coder-485B} with \texttt{gpt-oss\allowbreak-120B}.
In all two-model settings, we draw $M/2$ samples from each model and merge them into a set of $M$ programs, on which we compute uncertainty.

\subsection{RQ1. Correlation with Program Correctness}
\label{subsec:correlation_analysis}

To answer RQ1, we evaluate how these uncertainty scores correlate with program correctness. For each problem $x$, we evaluate all $M$ sampled programs on the official LiveCodeBench tests and obtain per-sample correctness scores $\{c_m\}_{m=1}^{M}$, where each score is defined as the proportion of hidden tests passed. We then summarize correctness by the mean score $\bar{c}(x) \coloneqq \frac{1}{M}\sum_{m=1}^{M} c_m$ and compute the Pearson correlation between $\bar{c}(x)$ and the corresponding uncertainty score $u(x)$ across problems. A stronger negative correlation indicates that higher uncertainty more faithfully signals lower functional correctness. The results are summarized in Table~\ref{tab:entropy_detailed}.

% To answer RQ1, we analyze how different uncertainty formulations correlate with program correctness under the default \textit{Func} clustering. For each problem $x$, we evaluate all $M$ sampled programs on the official LiveCodeBench tests and obtain per-sample correctness scores $\{c_m\}_{m=1}^{M}$, where each score is defined as the proportion of hidden tests passed. We then summarize correctness by the mean score $\bar{c}(x) \coloneqq \frac{1}{M}\sum_{m=1}^{M} c_m$ and compute the Pearson correlation between $\bar{c}(x)$ and the corresponding uncertainty score $u(x)$ across problems. A stronger negative correlation indicates that higher uncertainty more faithfully signals lower functional correctness.

% We compare five uncertainty scores. \textbf{PE} is the predictive entropy over token sequences introduced in Section~\ref{sec:bayesian_uq}. \textbf{SE} computes entropy after mapping outputs to semantic clusters (Section~\ref{sec:semantic_entropy}), and \textbf{DSE} is its discrete frequency-based counterpart. \textbf{ESE} and \textbf{EDSE} extend these semantic uncertainty measures to ensembles by aggregating samples across multiple models (Sections~\ref{sec:ese} and~\ref{sec:ens_dse}).

\textbf{Ensemble semantic entropy provides the strongest correctness signal.}
Across the comparable settings, \textbf{ESE} and \textbf{EDSE} consistently yield the strongest or nearly strongest correlations with correctness. Relative to the corresponding single-model baselines, the gains are especially clear for cross-family ensembles: the magnitude improves from -0.6738 to -0.7400 for \texttt{Qwen3-8B+GLM4-9B} and from -0.5961 to -0.7122 for \texttt{gpt-oss-20B+Qwen3-Coder-30B}. Within-family ensembling yields smaller but still consistent improvements, such as from -0.6852 to -0.7037 for \texttt{Qwen3-14B+Qwen3-8B} and from -0.6686 to -0.6998 for \texttt{Qwen3-Coder-30B+Qwen3-8B}. This pattern suggests that cross-model disagreement provides information about correctness beyond single-model self-consistency.

\textbf{Semantic entropy is much more correlated with program correctness than token-level predictive entropy.}
All semantic-entropy variants are strongly negatively correlated with correctness, typically around -0.6 to -0.74. In contrast, \textbf{PE} stays close to 0 and is even positive for some models, such as \texttt{Qwen3-8B} (0.0619), \texttt{Qwen3-14B} (0.2670), and \texttt{gpt-oss-20B} (0.1986). This gap is also reflected in statistical significance. For uncertainty methods based on semantic entropy, including \textbf{SE}, \textbf{DSE}, \textbf{ESE}, and \textbf{EDSE}, the $p$-values are consistently far below conventional thresholds, typically smaller than $10^{-40}$. In contrast, the correlation of \textbf{PE} with program correctness is much less stable, with some models yielding only weak significance or even insufficient significance (e.g., $p > 10^{-2}$). Taken together, these results show that token-level predictive entropy is not a reliable correctness signal for code generation, whereas semantic entropy provides a substantially more reliable correlate of program correctness.

\textbf{Ensembling matters more than refining single-model probability information.}
Notably, \textbf{SE} and \textbf{DSE} are almost indistinguishable across all backbones, while \textbf{ESE} and \textbf{EDSE} consistently improve over them. This suggests that the main gain comes from incorporating cross-model variation rather than from refining single-model probability information.

\begin{rqanswerbox}
\textbf{Answer to RQ1:} ESE and EDSE generally show stronger correlation with program correctness than single-model uncertainty methods, indicating that ensemble uncertainty provides a stronger correctness signal.
\end{rqanswerbox}

\subsection{RQ2. Effects of Different Clustering Methods}
\label{subsec:clustering_methods}

\begin{figure}[h]
    \centering
    \includegraphics[width=\linewidth]{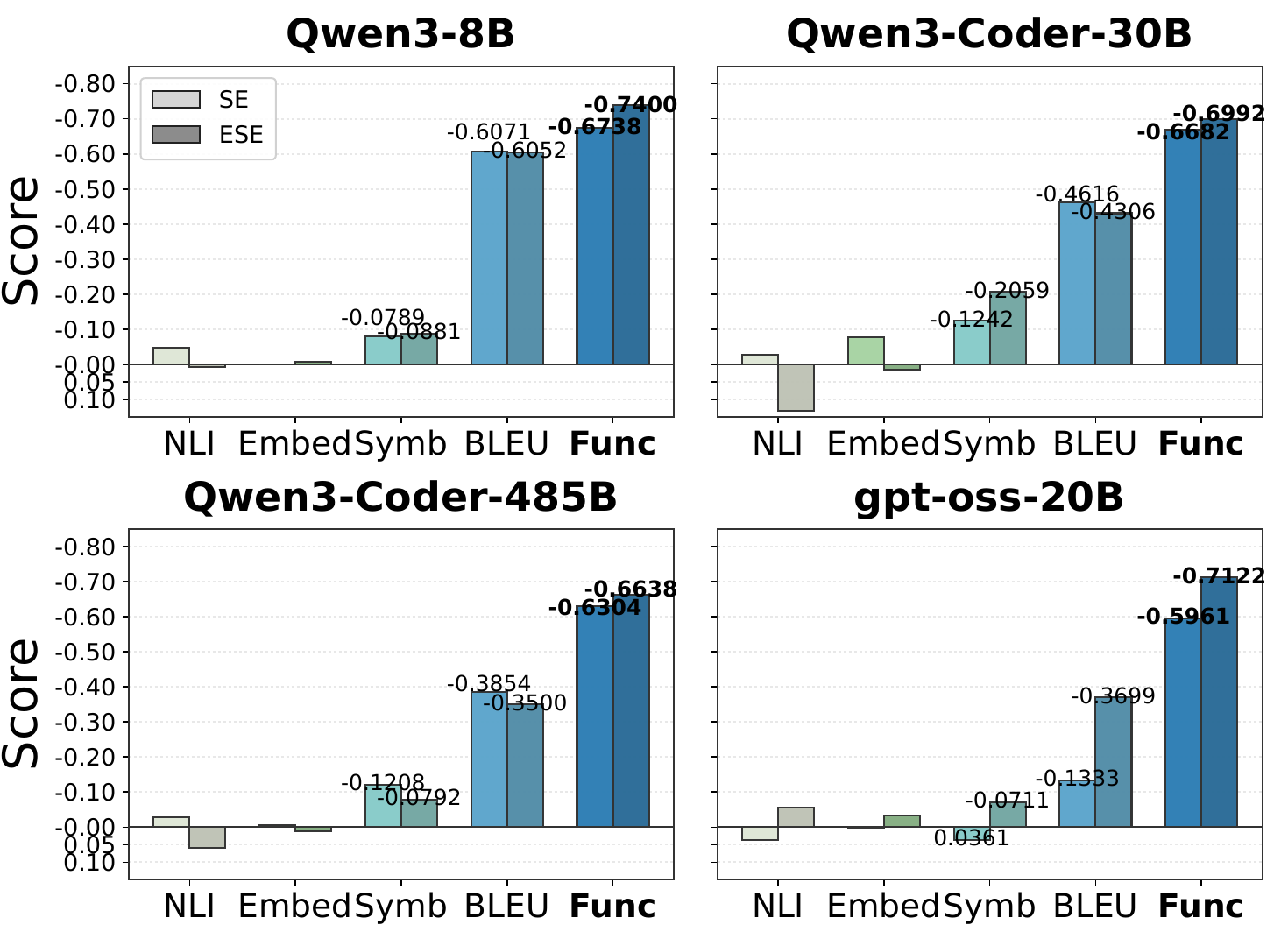}
    \caption{Pearson correlation coefficients between uncertainty and mean program correctness under different semantic clustering methods. For each method, the left bar denotes SE and the right bar denotes ESE. Results are reported for four representative backbones.}
    \label{fig:exp1_2}
\end{figure}

To answer RQ2, we study how the choice of clustering method affects the correlation between uncertainty and program correctness beyond the default \textit{Func} setting. We consider five clustering methods in this study, including \textit{NLI}, \textit{Embed}, \textit{BLEU}, \textit{Symbolic}, and \textit{Func}:
\begin{itemize}
    \item \textit{NLI}: uses \texttt{Deberta-v2-xlarge-mnli} to test bidirectional entailment between code pairs \cite{kuhn2023semanticuncertaintylinguisticinvariances};
    \item \textit{Embed}: clusters by distances in embedding space using \texttt{SFR-\allowbreak Embedding-Code-400M\_R};
    \item \textit{BLEU}: uses \texttt{CodeBLEU} \cite{ren2020codebleumethodautomaticevaluation} to measure program similarity via a mixture of AST, control-flow, and n-gram overlap;
    \item \textit{Symbolic}: applies \texttt{CrossHair} for symbolic execution \cite{sharma2025assessingcorrectnessllmbasedcode}, declaring two programs equivalent if no counterexample is found within the prescribed time budget. We set the per-path timeout to 5 seconds and the total timeout to 30 seconds;
    \item \textit{Func}: the behavior-based method introduced in Section~\ref{sec:semantic_clustering}, which clusters programs by their execution traces on test inputs.
\end{itemize}
We report the Pearson correlation between uncertainty measures and mean correctness under each clustering choice of four representative models in Figure~\ref{fig:exp1_2}.

\textbf{The choice of clustering method substantially affects uncertainty estimation.}
Figure~\ref{fig:exp1_2} shows that the clustering method has a clear impact on the correlation between uncertainty and correctness. Among the five methods, \textit{Func} consistently yields the strongest negative correlation across the representative backbones, while \textit{NLI}, \textit{Embed}, and \textit{Symbolic} are generally much weaker. \textit{BLEU} can partially capture algorithmic differences through control-structure overlap, yet it cannot reliably detect semantically critical changes that preserve surface control flow while altering functional behavior; consequently, uncertainty computed under \textit{BLEU} clustering remains weaker than the execution-grounded \textit{Func} variant. Finally, while symbolic analysis can in principle support path-complete reasoning, its practical use in our setting is limited: it typically requires manually specified preconditions to constrain the input space, and without such preconditions it may explore invalid inputs (e.g., negative values used as array lengths). Moreover, for complex programs it frequently exceeds the time budget and is then forced to conservatively declare equivalence. This trend suggests that, for code generation, uncertainty estimation benefits most from a semantic partition grounded in executable functional behavior rather than superficial similarity.

\input{table/acc}

\textbf{Ensembling improves correlation under nearly all clustering methods.}
Although the absolute quality of the clustering method varies, the effect of ensembling remains broadly consistent. In Figure~\ref{fig:exp1_2}, ESE achieves a stronger correlation than the corresponding SE result for almost every clustering method and model. This result shows that the benefit of ensemble-based uncertainty is not tied to a single clustering choice: once a semantic partition is induced, aggregating multiple models generally improves the correlation between uncertainty and program correctness.

\begin{rqanswerbox}
\textbf{Answer to RQ2:} Different clustering methods substantially affect the correlation between ESE and program correctness. Functional clustering yields the strongest correlation, while the correlation gain from ensembling remains consistent across nearly all clustering choices.
\end{rqanswerbox}

\subsection{RQ3. Performance on Selective Generation Task}
\label{subsec:abstention}

To answer RQ3, we evaluate whether the uncertainty signal can support accurate downstream decisions in selective generation tasks. In selective generation, the model should \emph{abstain} when it is likely to produce functionally incorrect code. To predict the correctness of generated code, We compute an uncertainty score $u(x)$ from the semantic clusters of the $M$ sampled programs on $\mathcal{T}^{gen}_x$. Given a threshold $\tau$, we \emph{accept} the prediction if $u(x) \le \tau$ and \emph{reject} it otherwise. Upon acceptance, we follow a majority-voting principle: we select the \emph{largest} semantic cluster and output the \emph{longest} program within that cluster as a deterministic selection rule. We then evaluate the selected program on the official LiveCodeBench tests and define it as \texttt{pass} if it succeeds (score $=1$) and \texttt{fail} otherwise.

We evaluate performance by the trade-off between \emph{False Positive Rate} (FPR), namely the fraction of \texttt{fail} cases that are incorrectly accepted, and \emph{True Positive Rate} (TPR), namely the fraction of \texttt{pass} cases that are correctly accepted. We report TPR under strict FPR constraints of 5\% and 10\% (TPR@5FPR, TPR@10FPR), together with the overall accept/reject classification accuracy (Acc.) at the 10\% FPR operating point. In addition to SE/DSE and their ensemble variants, we include \textbf{SE-NLI} as the original semantic entropy baseline that clusters samples using NLI-based entailment judgments. Results are shown in Table~\ref{tab:abs}.

\paragraph{\textbf{Ensemble-based uncertainty yields substantially better abstention decisions.}}
As shown in Table~\ref{tab:abs}, \textbf{ESE} substantially outperforms single-model uncertainty across all three evaluated backbones, and \textbf{EDSE} shows similarly strong performance. In particular, compared with \textbf{SE}, TPR@5FPR of \textbf{ESE} increases by 83\%, 172\%, and 49\% for \texttt{Qwen3-8B}, \texttt{Qwen3-Coder-30B}, and \texttt{gpt-oss-20B}, respectively, while the corresponding classification accuracy improves by 28.5\%, 53.2\%, and 37.8\%. \textbf{EDSE} attains the best results for \texttt{Qwen3-8B}, while remaining comparable to \textbf{ESE} on the other two backbones. These gains are consistent with the correlation results above: by incorporating cross-model disagreement, ensemble-based semantic entropy better detects confident but incorrect solutions that single-model self-consistency tends to miss, and therefore provides a more reliable signal for selective generation under strict false-positive constraints.

\begin{rqanswerbox}
\textbf{Answer to RQ3:} ESE supports more accurate selective-generation decisions than single-model baselines under strict false-positive constraints, showing that its correlation advantage transfers to downstream decision making.
\end{rqanswerbox}

%% file: table/auc_detailed.tex
% ===== Table 1: Main results (Func_gen only) =====
\begin{table*}[t]
    \centering
    \renewcommand\arraystretch{1.15}
    \caption{Pearson correlation coefficients and p-values between uncertainty and functional correctness on LiveCodeBench for each model (higher magnitude negative is better). We report \textbf{Predictive Entropy (PE)}, \textbf{Semantic Entropy (SE)}, and \textbf{Discrete Semantic Entropy (DSE)} under functional behavior clustering. Gray rows denote our proposed ensemble variants, Ensemble Semantic Entropy (ESE) and Ensemble Discrete Semantic Entropy (EDSE). \textbf{Bold} values indicate the best correlation coefficients, while \underline{underlined} values indicate the second-best results.}
    \label{tab:entropy_detailed}
    \scalebox{0.85}{
    \begin{tabular}{l|ccccccc}
    \toprule
    \textbf{Metric} & \textbf{GLM4-9B} & \textbf{Qwen3-8B} & \textbf{Qwen3-14B} & \textbf{Qwen3-Coder-30B} & \textbf{Qwen3-Coder-485B} & \textbf{gpt-oss-20B} & \textbf{gpt-oss-120B} \\
    \midrule
    
    PE  & -0.0899 ($4 \times 10^{-2}$) & 0.0619 ($1 \times 10^{-1}$) & 0.2670 ($3 \times 10^{-53}$) & -0.1066 ($1 \times 10^{-2}$) & -0.4183 ($5 \times 10^{-26}$) & 0.1986 ($6 \times 10^{-6}$) & -0.4320 ($5 \times 10^{-12}$) \\
    SE  & -0.7352 ($2 \times 10^{-101}$) & -0.6735 ($3 \times 10^{-81}$) & -0.6852 ($1 \times 10^{-75}$) & -0.6686 ($2 \times 10^{-67}$) & -0.6304 ($5 \times 10^{-58}$) & -0.5960 ($2 \times 10^{-50}$) & \textbf{-0.6751} ($1 \times 10^{-67}$) \\
    DSE & -0.7372 ($2 \times 10^{-102}$) & -0.6738 ($6 \times 10^{-81}$) & -0.6864 ($8 \times 10^{-76}$) & -0.6682 ($2 \times 10^{-67}$) & -0.6304 ($3 \times 10^{-58}$) & -0.5961 ($2 \times 10^{-50}$) & -0.6694 ($3 \times 10^{-67}$) \\
    \rowcolor{gray!10}
    \textit{ESE}  & -- & \underline{-0.7391} ($9 \times 10^{-90}$) & \underline{-0.7029} ($1 \times 10^{-77}$) & \textbf{-0.6998} ($1 \times 10^{-70}$) & \textbf{-0.6641} ($5 \times 10^{-63}$) & \underline{-0.7106} ($3 \times 10^{-80}$) & -0.6736 ($2 \times 10^{-43}$) \\
    \rowcolor{gray!10}
    \textit{EDSE} & -- & \textbf{-0.7400} ($2 \times 10^{-89}$) & \textbf{-0.7037} ($2 \times 10^{-77}$) & \underline{-0.6992} ($2 \times 10^{-70}$) & \underline{-0.6638} ($2 \times 10^{-64}$) & \textbf{-0.7122} ($9 \times 10^{-80}$) & \underline{-0.6750} ($6 \times 10^{-44}$) \\
    \bottomrule
    \end{tabular}
    }
    \end{table*}

%% file: table/acc.tex
\begin{table*}[t]
    \centering
    \renewcommand\arraystretch{1.15}
    \caption{Selective generation results on LiveCodeBench. We report TPR at fixed FPR thresholds (TPR@5FPR, TPR@10FPR) and the resulting classification accuracy (Acc.) at the 10\% FPR constraint. We use \textbf{SE-NLI}, \textbf{SE/DSE}, and the proposed \textbf{ESE/EDSE} as uncertainty scores to decide whether to abstain. \textbf{Bold} values indicate the highest accuracy or TPR.}
    \scalebox{0.9}{
    \begin{tabular}{l|ccc|ccc|ccc}
    \toprule
    \textbf{Method} 
    & \multicolumn{3}{c|}{\textbf{Qwen3-8B}} 
    & \multicolumn{3}{c|}{\textbf{Qwen3-Coder-30B}} 
    & \multicolumn{3}{c}{\textbf{gpt-oss-20B}} \\
    \cmidrule(lr){2-4} \cmidrule(lr){5-7} \cmidrule(lr){8-10}
    & Acc. & TPR@5FPR & TPR@10FPR
    & Acc. & TPR@5FPR & TPR@10FPR
    & Acc. & TPR@5FPR & TPR@10FPR \\
    \midrule
    SE-NLI    &0.4496  & 0.0500&  0.1001&0.3145  & 0.0460&  0.0920&0.1593  &0.0576  &0.1153 \\
    SE        &0.6420 & 0.3961&  0.6199& 0.4085 & 0.1797&  0.3600& 0.4671&0.4050  &0.8080 \\
    DSE       &0.6398 & 0.3920&  0.6139& 0.4078 & 0.1787&  0.3575& 0.4689&0.4070 &0.8106 \\
    ESE      &0.8247 & 0.7246&  0.7880& \textbf{0.6258} &\textbf{0.4889} &\textbf{0.7737}& \textbf{0.6435}&\textbf{0.6041}  &\textbf{0.8749} \\
    EDSE     &\textbf{0.8267} & \textbf{0.7283}&  \textbf{0.7980}& 0.5903 &0.4384 &0.7569& 0.6384&0.5983  &0.8668 \\
    \bottomrule
    \end{tabular}
    }
    \label{tab:abs}
    \end{table*}

%% file: section/tts.tex
\section{Application}
\label{sec:tts}

We instantiate the uncertainty measure from Section~\ref{sec:method} in a test-time scaling (TTS) pipeline for code generation, following the public/private test protocol used in programming contests and online judges.
Given a problem specification $x$, let $\mathcal{T}^{\text{pub}}_x$ and $\mathcal{T}^{\text{priv}}_x$ denote the public and hidden test suites, respectively.
A candidate program $y$ can be executed on $\mathcal{T}^{\text{pub}}_x$ to detect obvious misunderstandings or runtime errors, while correctness is ultimately determined by passing $\mathcal{T}^{\text{priv}}_x$.

Existing TTS frameworks primarily scale inference-time compute, such as parallel sampling and iterative refinement, for a fixed model, whereas model capability can also be improved by scaling model parameters \cite{yang2025testtimescalingllmssurvey,agarwal2025artscalingtesttimecompute}. In practice, scaling test-time compute alone often exhibits diminishing returns: substantially more computation may yield only marginal performance gains. Our goal is therefore to balance inference scaling and parameter scaling via a cascading mechanism: we start from cheaper models and escalate to stronger, more expensive models only when a decision module predicts that the current layer is unlikely to return a correct solution \cite{yue2024largelanguagemodelcascades}.

\subsection{\cas: Multi-layer Cascading Framework}

We propose an $H$-layer framework, \cas.
Each layer $\ell \in \{1,\dots,H\}$ is associated with an ensemble of $N_\ell$ code generators $\mathcal{M}^{(\ell)}=\{\theta^{(\ell)}_j\}_{j=1}^{N_\ell}$, where $\theta^{(\ell)}_j$ denotes the model instance used to generate the $j$-th candidate.
When all $\theta^{(\ell)}_j$ are identical, the layer reduces to standard single-model scaling.
On input $x$, the pipeline starts at $\ell=1$ and executes the following stages within each layer.

\textbf{(i) Parallel sampling.}
We draw $N_\ell$ candidate programs in parallel, producing $\mathcal{Y}^{(\ell)}=\{y^{(\ell)}_j\}_{j=1}^{N_\ell}$, where $y^{(\ell)}_j$ is sampled from $\theta^{(\ell)}_j$.

\textbf{(ii) Iterative public-test debugging.}
Each sampled program $y^{(\ell)}_j$ is allowed up to $D_\ell$ serial refinement steps, where $\theta^{(\ell)}_j$ uses feedback from executing $y^{(\ell)}_j$ on $\mathcal{T}^{\text{pub}}_x$ (e.g., failing cases or runtime errors) to produce an edited program.
We denote the resulting program by $\widetilde{y}^{(\ell)}_j$.
Let $\mathbb{I}[\widetilde{y}^{(\ell)}_j \text{ passes } \mathcal{T}^{\text{pub}}_x]$ be the indicator of public-test success, and define the \emph{public-pass ratio}
\begin{equation}
   \lambda_\ell(x) \coloneqq \frac{1}{N_\ell}\sum_{j=1}^{N_\ell}\mathbb{I}[\widetilde{y}^{(\ell)}_j \text{ passes } \mathcal{T}^{\text{pub}}_x]. 
\end{equation}
We retain only the passing candidates
$
\widetilde{\mathcal{Y}}^{(\ell)}_{\text{pass}}=\{\widetilde{y}^{(\ell)}_j: \widetilde{y}^{(\ell)}_j \text{ passes } \mathcal{T}^{\text{pub}}_x\}
$.

\textbf{(iii) Clustering-based decision and selection.}
Using the passing set $\widetilde{\mathcal{Y}}^{(\ell)}_{\text{pass}}$, we compute semantic clusters induced by the generated test set $\mathcal{T}^{\text{gen}}_x$ and evaluate the ensemble semantic uncertainty $\mathrm{ESE}_\ell(x)$.
If $\widetilde{\mathcal{Y}}^{(\ell)}_{\text{pass}}=\emptyset$ at a non-final layer, the current layer is rejected immediately and the pipeline proceeds to the next layer.
Otherwise, a \emph{cascade decision maker} compares a scalar score $S_\ell(x)$ against a threshold $\tau_\ell$ to decide whether to accept the current layer or escalate the problem to layer $\ell+1$.
If accepted, a selection rule returns a single program from $\widetilde{\mathcal{Y}}^{(\ell)}_{\text{pass}}$.
The final layer $\ell=H$ omits the rejection option, ensuring termination.

\subsection{Layerwise Cascade Decision Maker}

The decision maker integrates two complementary signals:
(1) \textbf{Plausibility}: represented by the public-pass ratio $\lambda_\ell$ (higher is better), indicating the models' overall ability to satisfy basic constraints.
(2) \textbf{Consensus}: represented by semantic uncertainty (lower is better). High agreement among samples from diverse models suggests reliability.

We define a normalized uncertainty $\widehat{u}_\ell(x)$ from $\mathrm{ESE}_\ell(x)$.
When the passing set induces at least two semantic clusters, we normalize by the maximum entropy under that partition:
$
\widehat{u}_\ell(x)=\frac{\mathrm{ESE}_\ell(x)}{\log |\widehat{\mathcal{C}}_\ell(x)|}.
$
When $|\widehat{\mathcal{C}}_\ell(x)|\le 1$, we set $\widehat{u}_\ell(x)=0$.
The cascade score is then defined as
\begin{equation}
\label{eq:cascade_score}
S_\ell(x) \coloneqq \alpha_\ell\,\lambda_\ell(x)-(1-\alpha_\ell)\,\widehat{u}_\ell(x),
\end{equation}
where $\alpha_\ell\in[0,1]$ balances the two terms.
The decision rule is $d_\ell(x)=\mathbb{I}[S_\ell(x)\ge \tau_\ell]$.
A lower $\tau_\ell$ encourages earlier acceptance at lower layers and lower cost, whereas a higher $\tau_\ell$ imposes stricter requirements on both public-test success and cross-sample agreement, thereby escalating uncertain problems to higher layers.

% This layerwise decision problem is closely related to the abstention setting studied in Section~\ref{subsec:abstention}: at each non-final layer, the system must decide whether the current candidates are reliable enough to accept or whether it should abstain and escalate.
% The difference is that cascading uses both the uncertainty signal and the public-pass ratio, since a practical TTS system must distinguish between semantically ambiguous candidates and candidates that already fail basic executable constraints.

\subsection{Selection via Semantic Clustering}

Upon acceptance at layer $\ell$ (or when $\ell=H$), the system must return a single program $\widehat{y}$.
We leverage the semantic partition $\widehat{\mathcal{C}}_\ell(x)$ computed during uncertainty estimation.
As at most one cluster corresponds to the correct behavior on $\mathcal{T}^{\text{gen}}_x$, we assume that the correct behavior is more likely to be reproduced consistently across generators than a spurious behavior and simply apply \emph{majority voting}:
\begin{equation}
c^\star_\ell(x) \coloneqq \arg\max_{c\in \widehat{\mathcal{C}}_\ell(x)} |c|.
\end{equation}
The final output is then sampled uniformly from the dominant cluster $c^\star_\ell(x)$.

Algorithm~\ref{alg:tts} summarizes the full \cas test-time scaling procedure.

\input{algorithms/tts}

\subsection{Evaluation}
\label{subsec:evaluation}

\begin{figure}
    \centering
    \includegraphics[width=\linewidth]{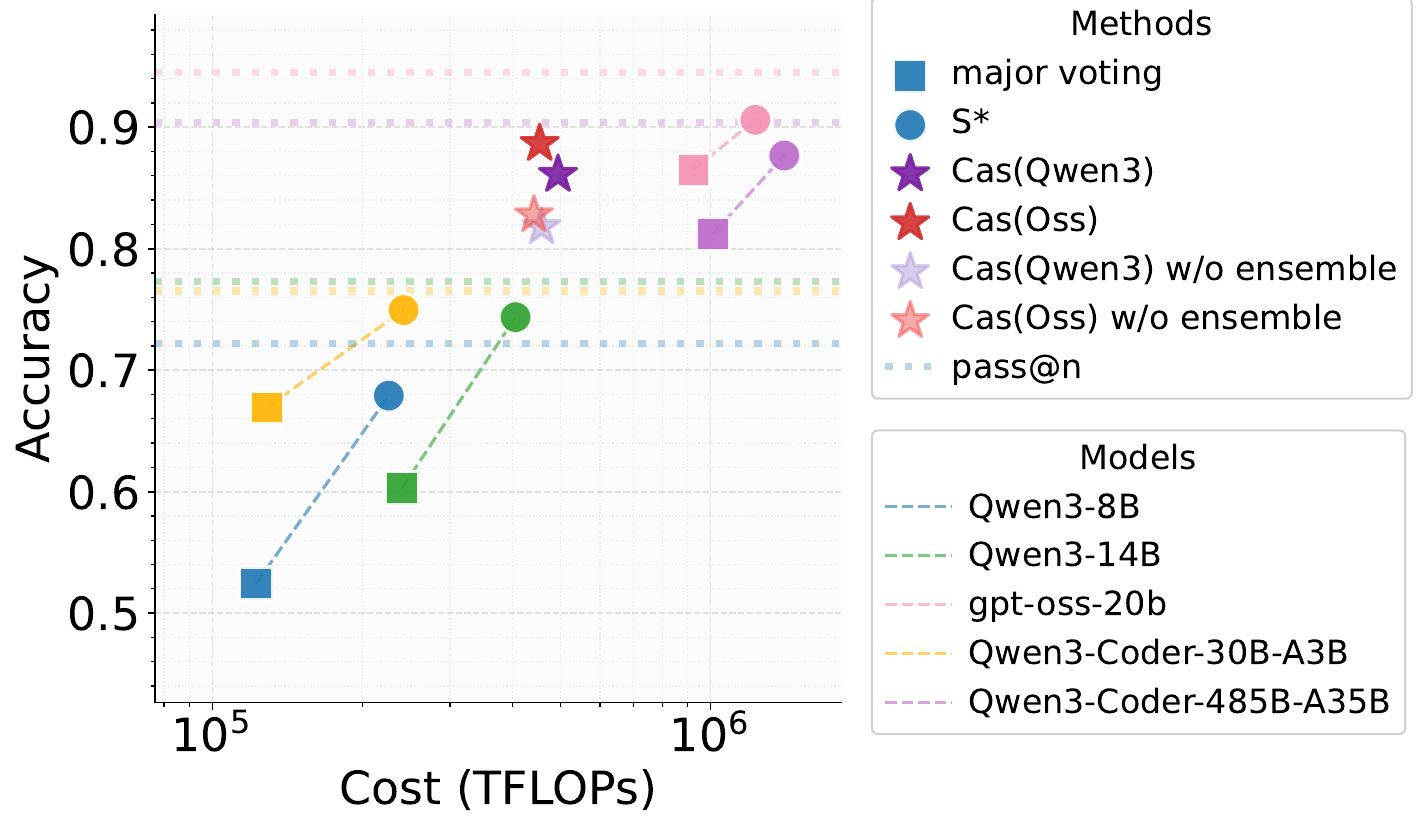}
    \vspace{-0.5cm}
    \caption{Accuracy-cost comparison on LiveCodeBench across four test-time scaling methods, Majority Voting, S*, \cas\ (w/o ensemble), and \cas, over five model configurations. Pass@20 is shown as the corresponding selection upper bound for each model. Methods in the upper-left region exhibit better accuracy-cost trade-offs.}
    \Description{Scatter plot of accuracy versus computational cost for multiple test-time scaling methods and model backbones; Cas variants lie near the high-accuracy, low-cost region compared with alternatives.}
    \label{fig:scale}
\end{figure}

\textbf{Experimental setting.}
We evaluate \cas ($H=2$) on two target backbones, \texttt{Qwen3-Coder-485B-A35B} and \texttt{gpt-oss-20B}, and compare it with single-model test-time scaling strategies.
The strongest baseline is \textbf{S*} \cite{li2025stesttimescaling}, which combines parallel sampling ($N=20$), iterative refinement ($D=3$), and best-of-$N$ selection on a single model.
In the cascading TTS framework \cas, the first layer is a lightweight ensemble of \texttt{Qwen3-8B} and \texttt{Qwen3-Coder-30B-A3B}, with 5 samples from each model.
The second layer uses 15 samples from a stronger target backbone, either \texttt{Qwen3-Coder-485B-A35B} or \texttt{gpt-oss-20B}; we denote the resulting framework by \texttt{Cas(Qwen)} and \texttt{Cas(Oss)}, respectively.
We set $\alpha_1=0.5$ and $\tau_1=0.3$.
For reference, we also report \textbf{Major Voting}, which applies semantic clustering and returns a program from the largest cluster, and \cas\ (w/o ensemble), which replaces the two-model ensemble in first layer with only \texttt{Qwen3-Coder-30B-A3B}. We report the  accuracy on LiveCodeBench and overall inference cost of each method in Figure~\ref{fig:scale}.

\textbf{\cas preserves near-best performance while substantially reducing inference cost, yielding the best overall accuracy-cost trade-off.}
As shown in Figure~\ref{fig:scale}, \cas lies closest to the upper-left frontier among the evaluated methods.
On \texttt{Qwen3-Coder\allowbreak-485B\allowbreak -A35B}, \texttt{Cas(Qwen)} reaches 86.11\% pass rate with cost 2.47, compared with 87.67\% and 7.03 for S*.
On \texttt{gpt-oss-20B}, \texttt{Cas(Oss)} reaches 88.65\% with cost 2.27, compared with 90.61\% and 6.15 for S*.
Thus, \cas reduces cost by 64.9\% and 63.1\%, respectively, while keeping the accuracy gap to S* within 1.56 and 1.96 percentage points.
Compared with simple Majority Voting, \cas improves both pass rate and inference cost, indicating that augmenting semantic-clustering-based selection with iterative public-test debugging improves candidate quality substantially, while incurring only a modest additional cost.
Moreover, replacing the first-layer ensemble with a single model causes a clear performance degradation, reducing pass rate by 4.31 and 5.87 percentage points for \texttt{Cas(Qwen)} and \texttt{Cas(Oss)}, respectively.
We next analyze these gains through the behavior of the cascade decision maker.

To further understand why \cas achieves high accuracy under low inference cost, we analyze the first-layer decision
$
d_1(x)=\mathbb{I}[S_1(x)\ge \tau_1],
$
where $S_1(x)=\alpha_1\lambda_1(x)-(1-\alpha_1)\widehat{u}_1(x)$ combines the public-pass ratio $\lambda_1(x)$ and the normalized uncertainty $\widehat{u}_1(x)$.
We first study the role of the threshold $\tau_1$, and then examine the contributions of the first-layer ensemble and the two decision signals.
In these analyses, we report two metrics. \textbf{Exit@L1} denotes the fraction of problems for which $d_1(x)=1$, i.e., the fraction of problems accepted at Layer~1 rather than escalated to Layer~2. \textbf{Pass@Exit} denotes the hidden-test pass rate of the selected programs among these accepted problems. These two metrics characterize the trade-off made by the cascade decision maker: increasing Exit@L1 reduces inference cost by accepting more problems at the first layer, whereas increasing Pass@Exit indicates that the selected programs returned by those first-layer decisions are more reliable.

We first study how the threshold $\tau_1$ controls the trade-off between acceptance rate and the quality of the selected program. We vary $\tau_1$ on \texttt{Cas(Qwen)} while keeping the other settings unchanged. Table~\ref{tab:thres} reports the resulting overall pass rate, average inference cost, Exit@L1, and Pass@Exit.

\input{table/cas_thres}

\textbf{$\tau_1=0.3$ provides the best balance between early-exit precision and inference cost.}
When $\tau_1$ is reduced to 0.20, the cascade decision maker accepts too many problems at Layer~1 while these selected programs are ultimately incorrect. Exit@L1 rises to 80.98\% and cost drops to 1.90, but Pass@Exit falls to 81.84\% and the final pass rate drops sharply to 79.21\%.
When $\tau_1$ increases beyond 0.30, the cascade decision maker becomes more conservative.
Pass@Exit further improves from 92.13\% to 93.31\% and 94.68\%, but many problems that could already be handled by the first layer are instead escalated to Layer~2, reducing Exit@L1 to 61.57\% and 55.29\% and increasing cost to 2.92 and 3.25, while the final pass rate improves only marginally to 86.50\% and 86.69\%.
Therefore, it is reasonable to set $\tau_1=0.3$, as it provides the best balance between pass rate and inference cost.

We next examine the contribution of each component in the cascade decision maker.
We consider three ablations: \cas\ (w/o ensemble), which replaces the first-layer ensemble with only \texttt{Qwen3-Coder\allowbreak-30B-A3B}; \cas\ (uncertainty only), which removes the public-pass term by setting $\alpha_1=0$, so that the decision score becomes
\begin{equation}
S_1^{\text{unc}}(x)=-\widehat{u}_1(x),
\end{equation}
with acceptance determined by a corresponding threshold on uncertainty; and \cas\ (public-pass only), which removes the uncertainty term by setting $\alpha_1=1$, so that
\begin{equation}
S_1^{\text{pub}}(x)=\lambda_1(x),
\end{equation}
with acceptance determined solely by the public-pass ratio.
Table~\ref{tab:cas_ablation} reports the resulting pass rate, cost, Exit@L1, and Pass@Exit on \texttt{Cas(Qwen)} and \texttt{Cas(Oss)}.

\input{table/cas_ablation}

\textbf{Utilizing ensemble semantic entropy as decision signal improves early-exit reliability by reducing false-positive exits.}
Relative to full \cas, \cas\ (w/o ensemble) reduces the final pass rate from 86.11\% to 81.80\% on \texttt{Qwen3-Coder-485B-A35B} and from 88.65\% to 82.78\% on \texttt{gpt-oss-20B}, while lowering Pass@Exit from 92.13\% to 86.98\% and increasing Exit@L1 from 69.80\% to 76.86\%.
In full \cas, the first-layer cascade decision maker evaluates $\mathrm{ESE}_1$ computed from the ensemble of \texttt{Qwen3-8B} and \texttt{Qwen3-Coder-30B-A3B}, whereas \cas\ (w/o ensemble) effectively relies on the semantic entropy of \texttt{Qwen3-Coder-30B-A3B} alone.
As discussed in Section~\ref{subsec:abstention}, single-model SE is more vulnerable to false-positive solutions: a model may generate candidate programs that are semantically consistent yet functionally incorrect, causing uncertainty to be underestimated and the problem to be accepted too early.
This effect explains why \cas\ (w/o ensemble) yields higher Exit@L1 but lower Pass@Exit.
By incorporating cross-model disagreement, ESE better identifies these false-positive solutions and escalates them to the stronger second layer, thereby improving the overall pass rate.

\textbf{Both Plausibility and Consensus signals in the cascade decision maker are necessary, as they capture different failure modes.}
\cas\ (uncertainty only) and \cas\ (public-pass only) both underperform the full cascade decision maker for different reasons.
\cas\ (uncertainty only), which relies only on uncertainty, reaches 79.01\% and 80.21\% final pass rate on the two backbones and lowers Pass@Exit to 74.40\%.
This result shows that low uncertainty alone is insufficient: for problems that the first layer cannot solve reliably, $\lambda_1(x)$ is often small, so the passing set $\widetilde{\mathcal{Y}}^{(1)}_{\text{pass}}$ is also small. Any apparent consensus in such a limited set is therefore unreliable, and $\widehat{u}_1(x)$ alone does not justify accepting the current layer.

\cas\ (public-pass only), which makes the Layer-1 acceptance decision using only the public-pass ratio, drops to 68.82\% and 69.93\% final pass rate, with Pass@Exit only 68.37\%, the worst result among all decision-maker variants.
This result shows that public tests alone are also insufficient: even when many candidates satisfy $\mathcal{T}^{\text{pub}}_x$, they may still occupy different semantic clusters on $\mathcal{T}^{\text{gen}}_x$ and disagree on hidden-test behavior, reflecting substantial uncertainty on these candidate programs.
Therefore, the construction of $S_1(x)$ is justified because $\lambda_1(x)$ and $\widehat{u}_1(x)$ capture different failure modes, and the current layer is reliable only when the public-pass ratio is sufficiently high and the semantic uncertainty is sufficiently low.

%% file: algorithms/tts.tex
\begin{algorithm}[h]
\caption{\cas \xspace Framework}
\label{alg:tts}
\begin{algorithmic}[1]
\REQUIRE Problem $x$, public tests $\mathcal{T}^{\mathrm{pub}}_x$, generated tests $\mathcal{T}^{\mathrm{gen}}_x$,
layer ensembles $\{\mathcal{M}^{(\ell)}\}_{\ell=1}^H$, 
sampling budgets $\{N_\ell\}$, debug budgets $\{D_\ell\}$,
cascade parameters $\{(\alpha_\ell,\tau_\ell)\}$.
\ENSURE Final program $\widehat{y}$

\FOR{$\ell = 1$ \ldots $H$}
    \STATE \textbf{Stage 1: Parallel Sampling}
    \STATE $\mathcal{Y}^{(\ell)} \gets \{ y^{(\ell)}_j \sim \theta^{(\ell)}_j(\cdot|x) \}_{j=1}^{N_\ell}$ where $\theta^{(\ell)}_j \in \mathcal{M}^{(\ell)}$

    \STATE \textbf{Stage 2: Public-test Debugging}
    \STATE $(\widetilde{\mathcal{Y}}^{(\ell)},\lambda_\ell) \gets 
    \textsc{DebugAndFilter}(\mathcal{Y}^{(\ell)}, \mathcal{T}^{\mathrm{pub}}_x, D_\ell)$
    
    \STATE \textbf{Stage 3: Clustering \& Uncertainty Estimation}
    \STATE $(\mathcal{C}_\ell, u_\ell) \gets 
    \textsc{UncertaintyEstimate}(\widetilde{\mathcal{Y}}^{(\ell)}, x)$

    \STATE \textbf{Stage 4: Cascade Decision \& Selection}
    \STATE $S_\ell \gets \alpha_\ell \lambda_\ell - (1-\alpha_\ell) u_\ell$
    \IF{$S_\ell \ge \tau_\ell$ \textbf{or} $\ell = H$}
        \STATE \textbf{return} \textsc{SelectMajority}$(\mathcal{C}_\ell)$
    \ENDIF
\ENDFOR
\end{algorithmic}
\end{algorithm}

%% file: table/cas_thres.tex
\begin{table}[!t]
    \centering
    \normalsize
    \setlength{\tabcolsep}{5.5pt}
    \renewcommand{\arraystretch}{1}
    \caption{Effect of the first-layer threshold $\tau_1$ on the trade-off between accuracy and inference cost (\texttt{Qwen3-Coder-485B}).}
    \begin{tabularx}{\columnwidth}{@{}
        >{\hsize=1.1\hsize\centering\arraybackslash}X
        |>{\hsize=0.9\hsize\centering\arraybackslash}X
         >{\hsize=0.9\hsize\centering\arraybackslash}X
         >{\hsize=0.9\hsize\centering\arraybackslash}X
         >{\hsize=0.9\hsize\centering\arraybackslash}X
    @{}}
    \toprule
    \textbf{$\tau_1$}
    & \textbf{Pass$\uparrow$} & \textbf{Cost$\downarrow$} & \textbf{Exit@L1$\uparrow$} & \textbf{Pass@Exit$\uparrow$}
    \\
    \midrule
    0.20
    & 79.21 & 1.90 & 80.98 & 81.84\\
    0.30
    & 86.11 & 2.47 & 69.80 & 92.13 \\
    0.40
    & 86.50 & 2.92 & 61.57 & 93.31 \\
    0.50
    & 86.69 & 3.25 & 55.29 & 94.68\\
    \bottomrule
    \end{tabularx}
    % \vspace{0.15cm}
    \label{tab:thres}
\end{table}

%% file: table/cas_ablation.tex
\begin{table*}[!t]
    \centering
    \renewcommand\arraystretch{1}
    \caption{Ablation analysis of the \cas cascade decision maker on the two models. Pass rate is reported in \%, cost is reported in $10^6$ TFLOPs, Exit@L1 denotes the fraction of instances accepted at Layer 1, and Pass@Exit denotes the hidden-test pass rate among these Layer-1 exits.}
    \resizebox{\textwidth}{!}{
    \begin{tabular}{l|cccc|cccc}
    \toprule
    \textbf{Method}
    & \multicolumn{4}{c|}{\textbf{Qwen3-Coder-485B}}
    & \multicolumn{4}{c}{\textbf{gpt-oss-20B}} \\
    \cmidrule(lr){2-5} \cmidrule(lr){6-9}
    & \textbf{Pass$\uparrow$} & \textbf{Cost$\downarrow$} & \textbf{Exit@L1$\uparrow$} & \textbf{Pass@Exit$\uparrow$}
    & \textbf{Pass$\uparrow$} & \textbf{Cost$\downarrow$} & \textbf{Exit@L1$\uparrow$} & \textbf{Pass@Exit$\uparrow$} \\
    \midrule
    \cas
    & 86.11 & 2.47 & 69.80 & 92.13
    & 88.65 & 2.27 & 69.80 & 92.13 \\
    \cas\ (w/o ensemble)
    & 81.80 & 2.29 & 76.86 & 86.98
    & 82.78 & 2.21 & 76.86 & 86.98 \\
    \cas\ (uncertainty only)
    & 79.01 & 2.28 & 73.52 & 74.40
    & 80.21 & 2.01 & 73.52 & 74.40 \\
    \cas\ (public-pass only)
    & 68.82 & 2.73 & 65.09 & 68.37
    & 69.93 & 2.13 & 65.09 & 68.37 \\
    \bottomrule
    \end{tabular}
    }
    \label{tab:cas_ablation}
\end{table*}

%% file: section/related_work.tex
\section{Related Work}
\textbf{Uncertainty Estimation in Code Generation.} 
Uncertainty estimation serves as a proxy for detecting hallucinations and predicting the correctness of responses from Large Language Models (LLMs). 
Concurrently, mainstream estimation techniques have evolved from token-level statistics to sampling-based consistency methods, most notably Semantic Entropy (SE) \cite{kuhn2023semanticuncertaintylinguisticinvariances,park2026efficientsemanticuncertaintyquantification,Farquhar2024,nguyen2025semanticentropyboostingllm}, which evaluates uncertainty by measuring entropy across clusters of semantically equivalent outputs. 
In the domain of code generation, such equivalence is rigorously defined by functional behavior; recent works utilize generated test cases \cite{ravuri2025eliminatinghallucinationinducederrorsllm,valentin2025incoherenceoraclelessmeasureerror} or symbolic execution \cite{sharma2025assessingcorrectnessllmbasedcode} to verify output consistency. 
Notably, \cite{dai2025reducinghallucinationsllmgeneratedcode} explores triangulation for open-ended problems, while our work mainly focuses on problems with a unique correct solution.

\noindent\textbf{Test-Time Scaling for Code Generation.} Test-time scaling (TTS) posits that increasing the computational budget during inference can significantly enhance model performance. In the domain of code generation, TTS strategies generally follow three main paradigms:
(i) \textbf{Extending Reasoning:} Extending reasoning length with techniques such as CoT \cite{wei2023chainofthoughtpromptingelicitsreasoning} and s1 \cite{muennighoff2025s1simpletesttimescaling} has proven effective. Specifically for code, Z1 \cite{yu2025z1efficienttesttimescaling} explores dynamic adjustments to reasoning length to optimize generation quality.
(ii) \textbf{Parallel Sampling and Voting:} Compared to single-pass generation, aggregating results from parallel sampling via Best-of-N selection \cite{huang2025bestofnbestthemcoverage} often yields significant performance gains. AlphaCode \cite{Li_2022, alphacode2} applies this strategy to code generation, leveraging large-scale parallel sampling and candidate selection to extend the capabilities of frontier models.
(iii) \textbf{Iterative Refinement:} The model's capacity for  self-correction \cite{madaan2023selfrefineiterativerefinementselffeedback} aligns naturally with execution-guided debugging. In this paradigm, models leverage original or self-generated test cases and utilize execution feedback to iteratively refine and debug solutions \cite{chen2022codetcodegenerationgenerated, chen2023teachinglargelanguagemodels, chen2025setsleveragingselfverificationselfcorrection}.
Recent works have focused on hybridizing these dimensions to push the Pareto frontier of performance and cost. S* \cite{li2025stesttimescaling} introduces a unified framework that augments parallel sampling with serial debugging, while GenCluster \cite{samadi2025scalingtesttimecomputeachieve} continuously scales test-time compute to IOI gold-level performance via parallel generation and behavioral clustering.

%% file: section/threats.tex
\section{Threats to Validity}
\label{sec:threats}

\noindent\textbf{Limited benchmark coverage.} Although we evaluate ESE and \cas on LiveCodeBench, which already provides a rigorous benchmark for recent competitive-programming tasks, this may limit the representativeness of our evaluation, future work can examine whether the same trends hold on broader real-world software engineering tasks.

\noindent\textbf{Limited model and ensemble coverage.} Although our experiments span three model families and include both within-family and cross-family two-model ensembles, this may limit the range of models and ensemble configurations, future work can study a wider range of backbones, larger ensembles, and alternative ensemble constructions.

\noindent\textbf{Limited semantic clustering method.} Although we compare several clustering methods and find functional clustering induced by $\mathcal{T}^{gen}_x$ to be strongest in our setting, this may introduce bias due to the reliance on generated tests, future work can explore stronger generated tests or hybrid semantic validation strategies.